\documentclass[aps,prl,reprint,amsmath,amssymb,longbibliography,nofootinbib,floatfix,superscriptaddress]{revtex4-2}

\usepackage{graphicx}
\usepackage{physics}
\usepackage{xcolor}
\usepackage{comment}
\usepackage{makecell}
\usepackage{hyperref}
\usepackage{bm}% bold math
\usepackage{letltxmacro}

\LetLtxMacro{\originaleqref}{\eqref}

\renewcommand{\d}{\partial}

\begin{document}

\title{Dynamic Water-Wave Tweezers 
%and Transport of Particles
}

\author{Jun Wang$^\#$}
\affiliation{Henan Key Laboratory of Quantum Materials and Quantum Energy, School of Quantum Information Future Technology, Henan University, Zhengzhou, 450046, China}
\affiliation{Institute of Quantum Materials and Physics, Henan Academy of Sciences, Zhengzhou, 450046, China}

\author{Shanhe Pang$^\#$}
\affiliation{Henan Key Laboratory of Quantum Materials and Quantum Energy, School of Quantum Information Future Technology, Henan University, Zhengzhou, 450046, China}

\author{Zhiyuan Che$^\#$}
\affiliation{State Key Laboratory of Surface Physics, Key Laboratory of Micro- and Nano-Photonic Structures (Ministry of Education), and Department of Physics, Fudan University, Yangpu District, Shanghai, 200433, China}
\affiliation{Institute of Acoustics, School of Physics Science and Engineering, Tongji University, 200092, Shanghai, China}

\author{Chang Liu}
\affiliation{Henan Key Laboratory of Quantum Materials and Quantum Energy, School of Quantum Information Future Technology, Henan University, Zhengzhou, 450046, China}
\affiliation{School of Electronic and Information Engineering, Anhui University, Hefei 230601, China}

\author{Wenzhe Liu}
\affiliation{State Key Laboratory of Surface Physics, Key Laboratory of Micro- and Nano-Photonic Structures (Ministry of Education), and Department of Physics, Fudan University, Yangpu District, Shanghai, 200433, China}

\author{Haoyu Xie}
\affiliation{Henan Key Laboratory of Quantum Materials and Quantum Energy, School of Quantum Information Future Technology, Henan University, Zhengzhou, 450046, China}

\author{Shipeng Li}
\affiliation{Henan Key Laboratory of Quantum Materials and Quantum Energy, School of Quantum Information Future Technology, Henan University, Zhengzhou, 450046, China}

\author{Zhongxia Du}
\affiliation{Henan Key Laboratory of Quantum Materials and Quantum Energy, School of Quantum Information Future Technology, Henan University, Zhengzhou, 450046, China}

\author{Xilai Hu}
\affiliation{Henan Key Laboratory of Quantum Materials and Quantum Energy, School of Quantum Information Future Technology, Henan University, Zhengzhou, 450046, China}

\author{Yanyong Li}
\affiliation{Henan Key Laboratory of Quantum Materials and Quantum Energy, School of Quantum Information Future Technology, Henan University, Zhengzhou, 450046, China}

\author{Bo Wang}
\email{bowang@henu.edu.cn}
\affiliation{Henan Key Laboratory of Quantum Materials and Quantum Energy, School of Quantum Information Future Technology, Henan University, Zhengzhou, 450046, China}
\affiliation{Institute of Quantum Materials and Physics, Henan Academy of Sciences, Zhengzhou, 450046, China}
\affiliation{Centre for Disruptive Photonic Technologies, School of Physical and Mathematical Sciences, Nanyang Technological University, Singapore 637371, Singapore}

\author{Lei Shi}
\email{lshi@fudan.edu.cn}
\affiliation{State Key Laboratory of Surface Physics, Key Laboratory of Micro- and Nano-Photonic Structures (Ministry of Education), and Department of Physics, Fudan University, Yangpu District, Shanghai, 200433, China}

\author{Konstantin Y. Bliokh}
\email{konstantin.bliokh@dipc.org}
\affiliation{Donostia International Physics Center (DIPC), Donostia-San Sebasti\'an 20018, Spain}
\affiliation{IKERBASQUE, Basque Foundation for Science, Bilbao 48009, Spain}
%\affiliation{Centre of Excellence ENSEMBLE3 Sp.~z o.o., 01-919 Warsaw, Poland}

\author{Yijie Shen}
\email{yijie.shen@ntu.edu.sg}
\affiliation{Centre for Disruptive Photonic Technologies, School of Physical and Mathematical Sciences, Nanyang Technological University, Singapore 637371, Singapore}
\affiliation{School of Electrical and Electronic Engineering, Nanyang Technological University, Singapore 639798, Singapore}

%TC:ignore
\begin{abstract}
Following a recent demonstration of stable trapping of floating particles by stationary (monochromatic) structured water waves [{\it Nature} {\bf 638}, 394 (2025)], we report dynamic water-wave tweezers that enable {\it controllable transport} of the trapped particle along an arbitrary trajectory on the water surface. {Furthermore, we demonstrate simultaneous transport of two trapped particles along different trajectories.}
We employ a triangular lattice formed by the interference of three plane waves, which can trap particles, depending on {the wave frequency and particle parameters}, either at intensity maxima or at intensity zeros (vortices). By introducing small frequency detunings of the interfering waves, we control 2D motion of the lattice and trapped particles. 
This approach is robust and effective over a relatively broad range of particle sizes and wave frequencies, offering remarkable new possibilities for noncontact manipulation of floating (e.g., biological and soft-matter) objects in fluidic environments.
\end{abstract}
%TC:endignore

\maketitle

\def\thefootnote{\#}\footnotetext{These authors contributed equally to this work.}

%%%%%%%%%%%%%%%%%%%%%%%%%%%%%%%%%%%%%%%%%
{\it Introduction.---}
%%%%%%%%%%%%%%%%%%%%%%%%%%%%%%%%%%%%%%%%%
Optical and acoustic tweezers are invaluable tools for the manipulation of various particles, ranging from individual atoms to living organisms  \cite{Ashkin_book, Molloy2002CP, Grier2003N, Bustamante2021NR, Gieseler2021AOP, Volpe2023JPP,  Ozcelik2018NM, Meng2019JPD, Baudoin2020AR, Dholakia2020NRP, Toftul2025arxiv, xie2026topological}.
They have found numerous applications, from quantum simulators and volumetric displays to optomechanical devices and biological cell sorting. Wave-based tweezers operate by trapping particles at maxima or minima of the wave intensity, as first demonstrated by Ashkin for optical laser fields \cite{Ashkin1970}. 
Together, optical and acoustic tweezers allow trapping and manipulation of particles with sizes spanning from $\sim 100\;$nm to $\sim 1\;$cm \cite{Dholakia2020NRP, Toftul2025arxiv}. 

{At the same time, stable trapping of floating particles using structured {\it water surface waves} with inhomogeneous (yet stationary) intensity landscapes was demonstrated in Refs.~\cite{Falkovich2005N, Denissenko2006PRL, Lukaschuk2007EPJ} and has recently been employed for particle manipulation \cite{Wang2025N}.} Here, we extend this idea by using {\it time-varying} water-wave landscapes that enable both trapping and {\it controllable transport} of floating particles along arbitrary trajectories on the surface. 
We execute the manipulation of a single particle, as well as the simultaneous independent manipulation of two particles.

{Our results provide the 2D water-wave counterparts of acoustic and optical 1D ``conveyors'' \cite{Whitworth1991U, Ruffner2012PRL_II, Li2021SA} and 2D dynamical trapping landscapes \cite{Courtney2012PRSA, Courtney2014APL, Morrell2024PRE}.}
These {proof-of-principle demonstrations} opens a pathway toward effective, noninvasive manipulation of floating particles, including biological and soft-matter objects, in the size range where optical and acoustic tweezers become inefficient: from millimeter up to meter scales.

%FFFFFFFFFFFFFFFFFFFFFFFFFFFFFFFFFFFFFFFFFFFFFFFFFFFFFFFFFFF
\begin{figure}[t]
\centering
\includegraphics[width=0.9\linewidth]{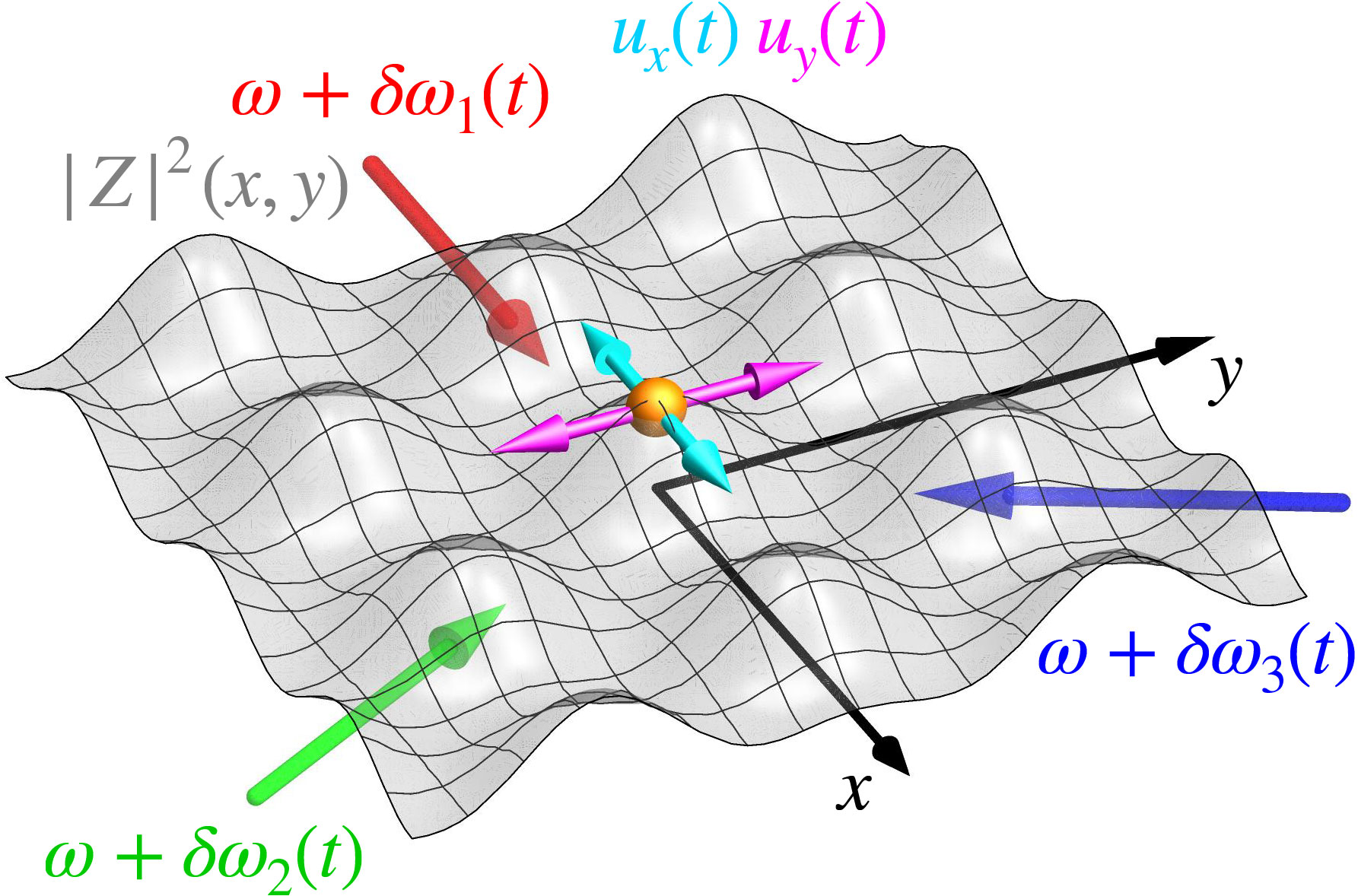}
\caption{Schematics of the interference of three plane waves with time-varying frequencies $\omega+\delta\omega_i(t)$, $|\delta\omega_i| \ll \omega$. This produces a near-triangular lattice in the elevation wavefield $Z(x,y)$ (its intensity is shown by the gray surface), which traps and transports a floating particle with a local velocity ${\bf u}(t)$.}
\label{Fig_1}
\end{figure}
%FFFFFFFFFFFFFFFFFFFFFFFFFFFFFFFFFFFFFFFFFFFFFFFFFFFFFFFFFFF

%%%%%%%%%%%%%%%%%%%%%%%%%%%%%%%%%%%%%%%%%
{\it General approach.---}
%%%%%%%%%%%%%%%%%%%%%%%%%%%%%%%%%%%%%%%%%
We employ the setup described in Refs.~\cite{Smirnova2024PRL, Wang2025N}, based on the interference of three plane water waves of equal amplitude, propagating with azimuthal directions $\varphi_i=(0,2\pi/3,4\pi/3)$, $i={1,2,3}$, see Fig.~\ref{Fig_1}. When these waves have the same frequency $\omega$, their interference produces a triangular lattice of intensity maxima and nodal points (i.e., {\it wave vortices} \cite{Nye1974, Dennis2009PO, Smirnova2024PRL, Wang2025N}) in the complex water-surface elevation field $Z({\bf r})$, ${\bf r} \equiv (x,y)$. The real elevation field is given by ${\mathcal Z ({\bf r},t)} = {\rm Re}\!\left[Z({\bf r}) e^{-i\omega t}\right]$. As demonstrated in Refs.~\cite{Falkovich2005N, Denissenko2006PRL, Lukaschuk2007EPJ, Wang2025N}, subwavelength-sized floating particles can be stably trapped either at the intensity maxima or at the zeros of the field. 

When the frequency of one of the waves is slightly detuned from $\omega$, the vortex lattice begins to move in the direction of this wave \cite{Smirnova2024PRL}. 
The complex wavefield becomes time-dependent, $Z({\bf r},t)$, and the moving vortices in such lattice are known as {\it spatiotemporal vortices} \cite{Sukhorukov2005, Bliokh2012, Jhajj2016, Chong2020, Zhang2023NC, Che2024PRL, Martin2025NP}. We write the wave frequencies as $\omega_{i} = \omega + \delta\omega_i$, $|\delta\omega_i|\ll \omega$, and the corresponding wavevectors as ${\bf k}_i =k_i {\bf e}_i = k_i(\cos\varphi_i, \sin\varphi_i)$. Here, $k_i\simeq k + \delta\omega_i/v_g$, where $k$ is the central wavenumber associated with frequency $\omega$, and $v_g = \d \omega/\d k$ is the wave's group velocity. 

The small variations in wavenumbers $k_i$ only slightly deform the lattice, which can be neglected in the leading-order approximation. In turn, the frequency detunings $\delta\omega_i$ produce a global drift of the lattice with velocity ${\bf u} \simeq (1/3k)\Sigma_{i} \delta\omega_i {\bf e}_i$. Explicitly, the Cartesian components of the velocity read:
\begin{equation}
\label{eq:velocity}
u_x \simeq \frac{1}{3k}\!\left[ 2\delta\omega_1-(\delta\omega_2 + \delta\omega_3)\right], ~
u_y \simeq \frac{1}{\sqrt{3}k}(\delta\omega_2 - \delta\omega_3)\,.
%{\bf k}_3 = k_2\left(-\frac{1}{2},\frac{\sqrt{3}}{2}\right)\,,
\end{equation}

By varying the detunings in time, $\delta\omega_i = \delta\omega_i (t)$, the lattice velocity ${\bf u}(t)$ can be dynamically controlled, producing motion along  trajectory ${\bf R}(t) = {\bf R}(0) + \int_0^t {\bf u}(t') dt'$. 
It is sufficient to vary only two of the three frequencies to generate motion in an arbitrary direction in the $(x,y)$ plane. In our experiments, we varied all three frequencies to facilitate control of the particle trajectories. 
{Moreover, by varying the central frequency $\omega$ and hence the wavelength, one can dynamically {\it scale} the lattice. This provides an additional degree of freedom enabling the independent manipulation of two particles trapped at different lattice sites.}

\begin{figure}[t]
\centering
\includegraphics[width=\linewidth]{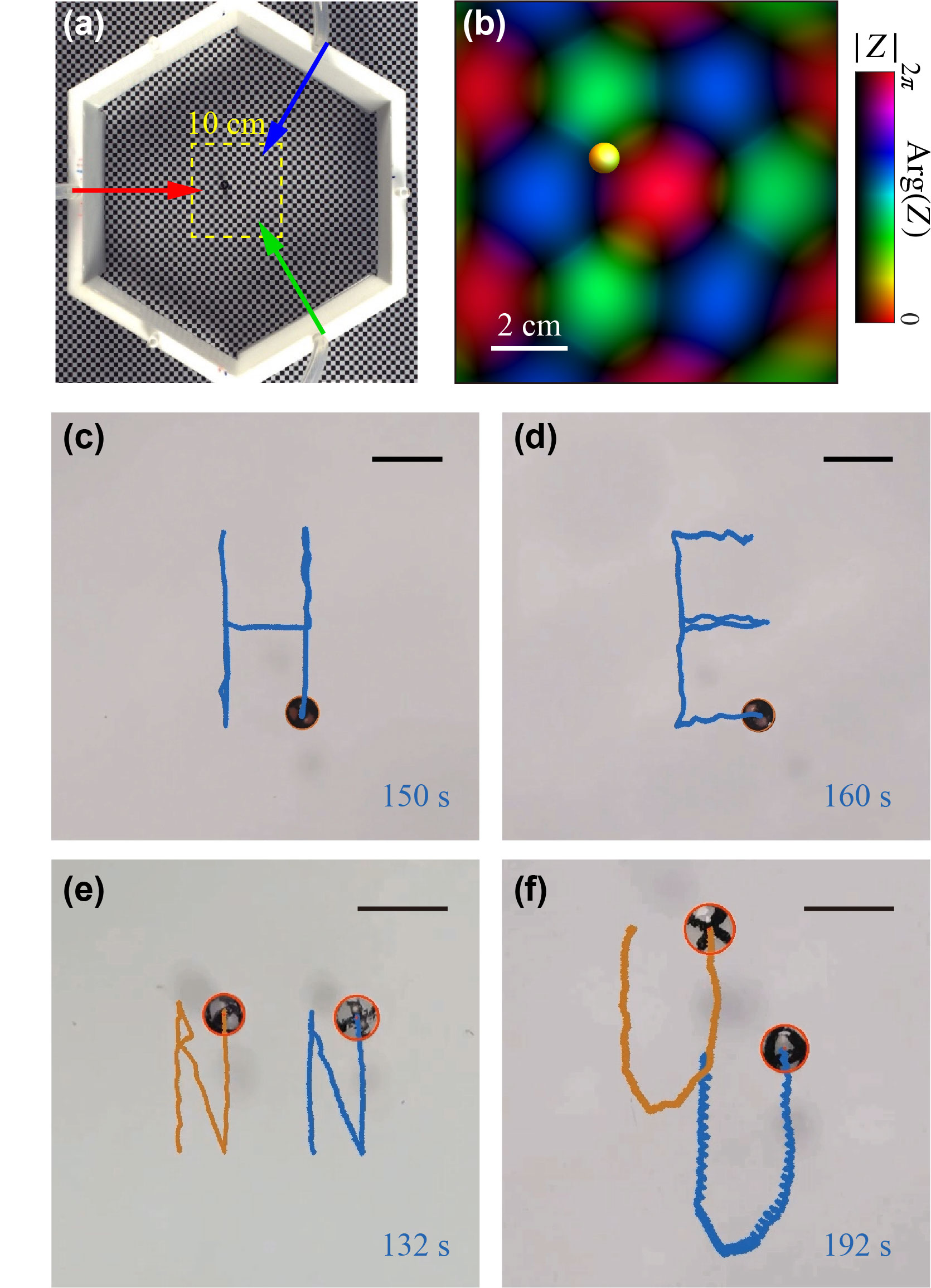}
\caption{(a) Top view of the experimental setup. (b) Measured complex field $Z({\bf r})$ within the dashed yellow square in (a) for stationary three-wave interference with central frequency $f = 6\;$Hz.
Brightness and hue colors represent the field amplitude and phase, respectively.
A particle of diameter $d=9.5\;$mm, trapped at a field zero (vortex), is shown schematically. (c--f) Transport of trapped particles achieved by modulating the frequencies of the interfering waves; see also Supplemental Movies 1--4, Fig.~\ref{Fig_freq}, and Supplemental Fig.~S1 \cite{SM1}. All scalebars throughout the paper correspond to 2~cm.}
\label{Fig_exp}
\end{figure}

%%%%%%%%%%%%%%%%%%%%%%%%%%%%%%%%%%%%%%%%%
{\it Main experiments.---}
%%%%%%%%%%%%%%%%%%%%%%%%%%%%%%%%%%%%%%%%%
The experiments were performed in a $1.2\times1.2\,{\rm m}^2$ water tank with depth $h = 2.5\;$cm, see Fig.~\ref{Fig_exp}(a). 
Similarly to Ref.~\cite{Wang2025N}, the interference field was generated inside a hexagonal structure with side length 16~cm, where three alternate sides acted as independent plane-wave sources, while the opposite sides served as open boundaries (raised above the water surface). 
The entire structure was surrounded by sponge absorbers to suppress wave reflections. 
The source sides were connected to speakers controlled by a multichannel sound card interfaced with a computer. This produced a three-wave interference field in the central $10\times 10\;{\rm cm}^2$ region. 
{In all experiments, the maximum wave amplitude in the three-wave interference region was $|Z|_{\rm max} \simeq 1\!-\!1.5\;$mm.}

In the first experiment, we used central frequency $f = \omega/2\pi=6\,{\rm Hz}$, corresponding to wavelength {$\lambda = 2\pi/k \simeq 4.8\,{\rm cm}$}. The water-surface elevation ${\mathcal Z ({\bf r},t)}$ was measured using fast checkerboard demodulation (FCD) \cite{Wildeman2018}, and the corresponding complex field $Z({\bf r})$ was reconstructed via the Hilbert transform, see Fig.~\ref{Fig_exp}(b). 
Under these conditions, spherical polyethylene particles with density $\rho \simeq 0.9\;{\rm g/cm^3}$ and diameter $d \simeq9.5\;$mm {(which corresponds to the Rayliegh wave-particle interaction regime with $d/\lambda \simeq 0.2$ \cite{Toftul2025arxiv})} were stably trapped at vortices, i.e., zeros of the $Z({\bf r})$ field, as shown in Fig.~\ref{Fig_exp}(b). 

%FFFFFFFFFFFFFFFFFFFFFFFFFFFFFFFFFFFFFFFFFFFFFFFFFFFFFFFFFFF
%FFFFFFFFFFFFFFFFFFFFFFFFFFFFFFFFFFFFFFFFFFFFFFFFFFFFFFFFFFF

Then, we introduced small frequency detunings within the range $|\delta f_i|\leq 0.05\,{\rm Hz}$. By controlling the direction of motion according to Eq.~\eqref{eq:velocity}, we were able to transport the particle along arbitrary prescribed trajectories. As an illustration, Figs.~\ref{Fig_exp}(c--f) show particle transport along paths forming the letters ``HENU'' (see also Supplemental Movies~1--4). Moreover, for the letters ``N'' and ``U'', two particles were simultaneously trapped and transported at different vortex sites of the lattice, see Figs.~\ref{Fig_exp}(e,f). Additional examples, corresponding to the letters ``NTU'' and ``DIPC'', are presented in the Supplemental Fig.~S2 \cite{SM1} and 
Supplemental Movies~5--11.

\begin{figure}[t]
\centering
\includegraphics[width=\linewidth]{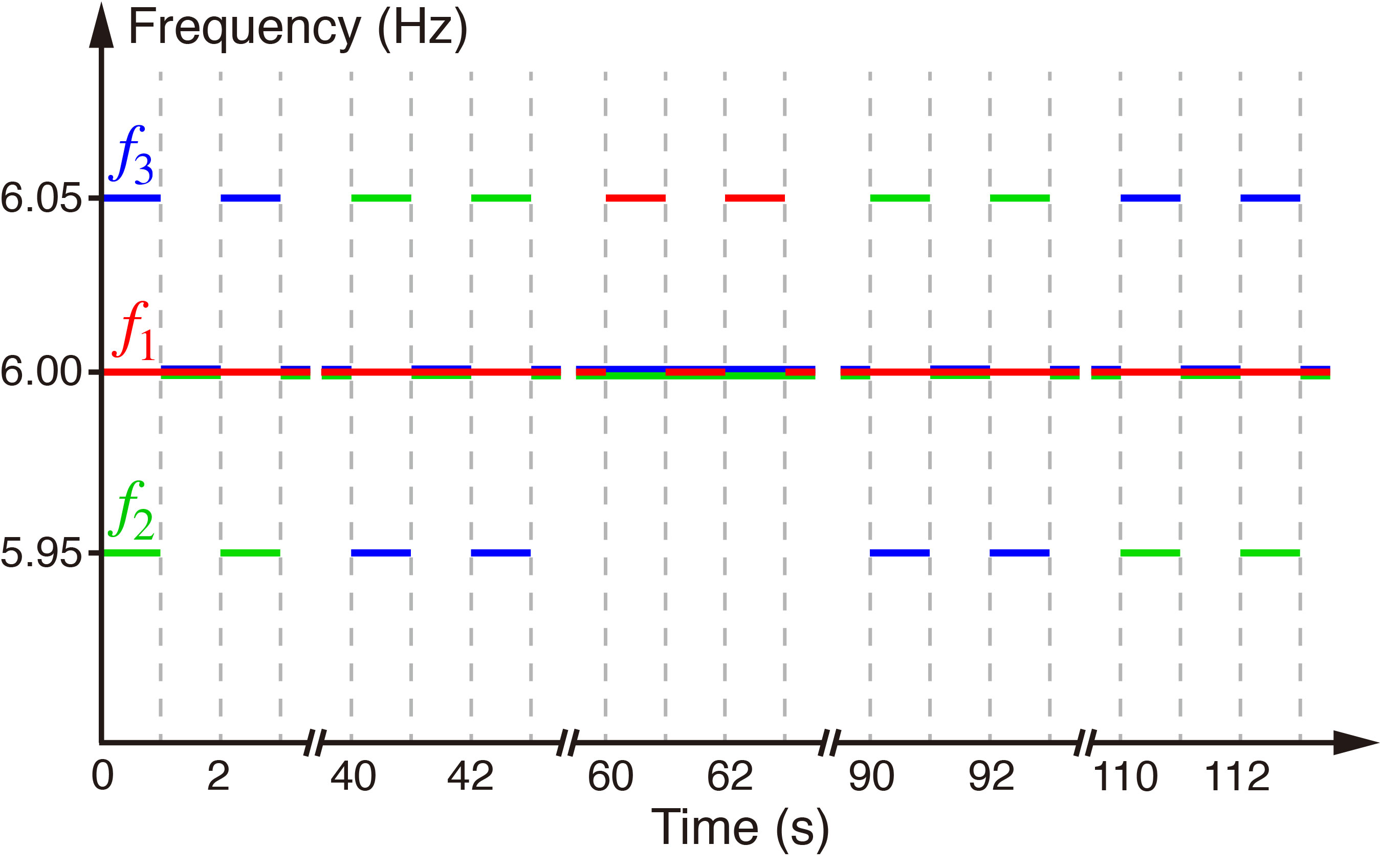}
\caption{Temporal dependencies of the frequencies $f_i(t) = \omega_i(t)/2\pi$ corresponding to particle transport along the letter-``H'' trajectory shown in Fig.~\ref{Fig_exp}(c) and Supplemental Movie~1 (see also Supplemental Fig.~S1 \cite{SM1}).}
\label{Fig_freq}
\end{figure}

{To ensure stable trapping of the particle throughout the transport process}, we implemented a piecewise frequency-modulation scheme. Specifically, the required frequency detunings $\delta\omega_i$ were applied during every odd second, while the lattice was stabilized by setting $\delta\omega_i=0$ during every even second. As an example, Fig.~\ref{Fig_freq} and Supplemental Fig.~S1 \cite{SM1} show the temporal dependencies $f_i(t)=\omega_i(t)/2\pi$ corresponding to particle transports along the ``HENU'' trajectories shown in Figs.~\ref{Fig_exp}(c--f). 
{Without these stabilization intervals, the dynamical trapping becomes less stable, particularly when the trajectory involves sharp turns (see Supplemental Fig.~S3 \cite{SM1}).}

%FFFFFFFFFFFFFFFFFFFFFFFFFFFFFFFFFFFFFFFFFFFFFFFFFFFFFFFFFFF
%FFFFFFFFFFFFFFFFFFFFFFFFFFFFFFFFFFFFFFFFFFFFFFFFFFFFFFFFFFF

\begin{figure}[t]
\centering
\includegraphics[width=0.7\linewidth]{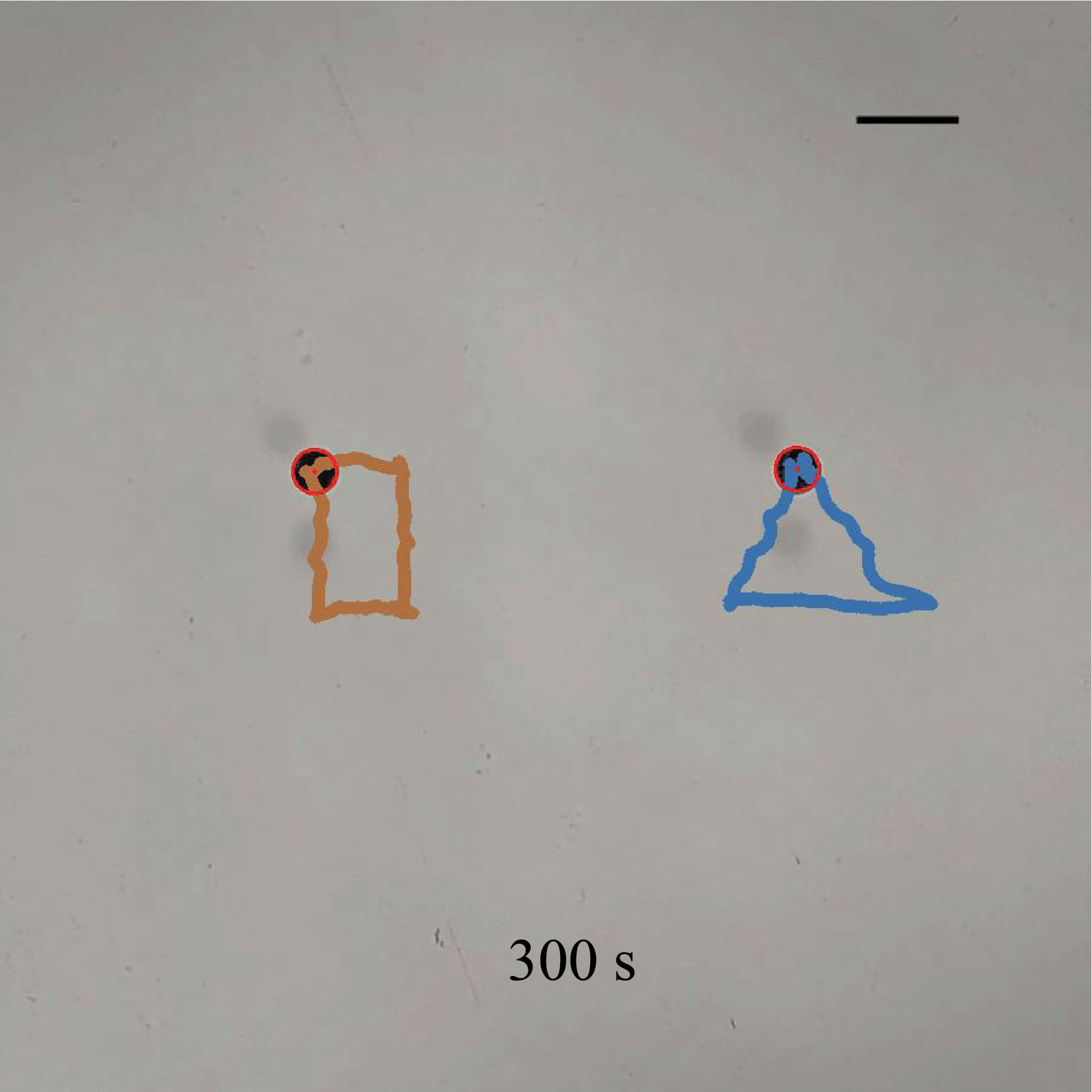}
\caption{Simultaneous transport of two particles, trapped at different lattice sites, along square and triangular trajectories. The distance between the particles is controlled by dynamically scaling the lattice through variations of the central frequency, $\omega (t)$. See also Appendix A and Supplemental Movies 12 and 13.}
\label{Fig_two_particles}
\end{figure}

{In the second experiment, we executed the simultaneous transport of two trapped particles along {\it different} trajectories. Their collective motion was controlled by the time-varying phases of the three waves, $\phi_i(t)$, which are effectively equivalent to the frequency detunings $\delta\omega_i(t) = d\phi_i/dt$). 
In turn, the relative distance between the particles was controlled by varying the central frequency, $\omega(t)$, thereby changing the wavelength $\lambda (t)$ and dynamically {\it scaling} the triangular lattice (see Fig.~\ref{Fig_trap} below). 
By combining these two types of motion, one can realize distinct trajectories for two particles trapped at different lattice sites.
Figure~\ref{Fig_two_particles} and Supplemental Movie 12 show an example of two-particle transport along square and triangular trajectories. 
The corresponding temporal dependencies $\phi_i(t)$ and $\omega(t)$, together with the evolution of the lattice, are provided in Appendix A and Supplemental Movie 13.}

%FFFFFFFFFFFFFFFFFFFFFFFFFFFFFFFFFFFFFFFFFFFFFFFFFFFFFFFFFFF
%FFFFFFFFFFFFFFFFFFFFFFFFFFFFFFFFFFFFFFFFFFFFFFFFFFFFFFFFFFF

%%%%%%%%%%%%%%%%%%%%%%%%%%%%%%%%%%%%%%%%%
{\it Trapping and stability tests.---}
%%%%%%%%%%%%%%%%%%%%%%%%%%%%%%%%%%%%%%%%%
We investigated various regimes and parameter ranges for stable trapping and transport of floating particles. First, we examined the trapping of a particle with diameter $d=9.5\;$mm in monochromatic three-wave interference fields at different frequencies $f$. The results are presented in Fig.~\ref{Fig_trap} and Fig.~\ref{Fig_exp}(b). At lower frequencies, $f = 4\!-\!5\;$Hz, the particle is stably trapped at an {\it intensity maximum} (anti-nodal point) of $Z({\bf r})$, whereas at higher frequencies, $f = 6\!-\!7\;$Hz, it is trapped at a {\it vortex} (nodal point) of $Z({\bf r})$. Both trapping regimes are suitable for particle transport in frequency-modulated fields. The transition between these regimes occurs near $f=5.5\;$Hz, where the particle undergoes orbital rotation around the vortex center, similar to the behavior observed in \cite{Wang2025N}.

We emphasize that the two observed trapping regimes are controlled by the wave frequency for the {\it same} particle. This contrasts with the nodal and anti-nodal trapping of {\it different} particles with hydrophilic and hydrophobic properties reported in Refs.~\cite{Falkovich2005N, Denissenko2006PRL, Lukaschuk2007EPJ}. Remarkably, we found that the transition between the two regimes occurs near the eigen-frequency $f_0$ of vertical oscillations of the particle floating on a still water surface. Neglecting capillary effects, $f_0=\sqrt{6g \rho_0 \delta_0(d - \delta_0)/(\rho d^3)} \, /2\pi \simeq 5.24\,$Hz (see Appendix B), where $\rho_0$ is the water density, $g$ is the gravitational acceleration, and $\delta_0$ is the depth of the submerged part of the particle. 
The sign flip of the trapping gradient force near the resonant frequency is also typical for optical and acoustic trapping \cite{Toftul2025arxiv}; 
it is accompanied by a reversal of the frequency‑dependent oscillatory response of the particle.
Such a transition between different trapping regimes was not observed in \cite{Falkovich2005N, Denissenko2006PRL, Lukaschuk2007EPJ} because those works used the low-frequency regime: $f \ll f_0$ \cite{Lukaschuk2007EPJ}.

\begin{figure}[t]
\centering
\includegraphics[width=\linewidth]{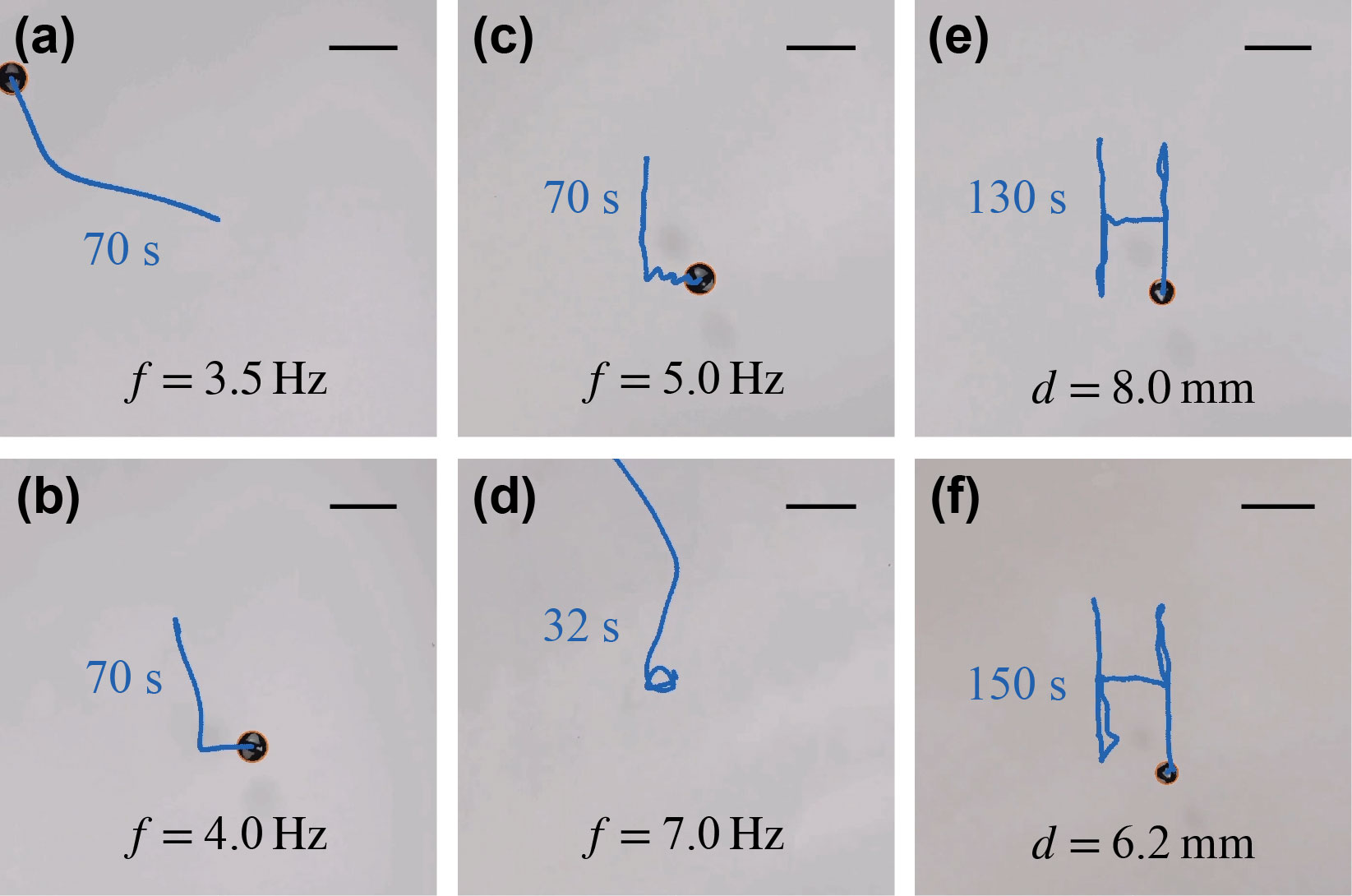}
\caption{Stability of particle transport at different parameter values. (a--d) Particle of diameter $d=9.5\;$mm in frequency-modulated fields at different central frequencies $f=3.5\!-\!7\;$Hz (see also Supplemental Movies~14--21 and Supplemental Fig.~S4 \cite{SM1}). (a,d) Unstable regimes. (b,c) Stable trapping at an intensity maximum and transport along an L-shaped path. (e,f) Controllable transport in a field with $f=6\;$Hz, similar to Fig.~\ref{Fig_exp}(c), but for particles with diameters $d=8.0$ and 6.2~mm (see also Supplemental Movies~22 and 23).}
\label{Fig_stability}
\end{figure}

Next, Figs.~\ref{Fig_stability}(a--d) present experiments on transport of particles of diameter $d=9.5\;$mm in frequency-modulated fields with different central frequencies (see also Supplemental Movies~14--21 and Supplemental Fig.~S4 \cite{SM1}). These results show that trapping and transport become unstable outside the frequency range $f = 4\!-\!6.5\;$Hz, see Figs.~\ref{Fig_stability}(a,d). Within this range, transport remains stable regardless of the trapping regime. In particular, Figs.~\ref{Fig_stability}(b,c) demonstrate controllable transport along an L-shaped trajectory using trapping at a wave-intensity maximum.
The unstable behavior observed at lower and higher frequencies may result from the breakdown of the ponderomotive-force regime, which relies on averaging over the wave-oscillation period $f^{-1}$, as well as from approaching the regime where the field-gradient scale $k^{-1}$ becomes comparable to the paricle size, see Fig.~\ref{Fig_trap}(f).

Finally, we explored transport of particles with different sizes in a frequency-modulated field with central frequency $f=6\;$Hz. Figures~\ref{Fig_stability}(e,f) and Fig.~\ref{Fig_exp}(c--f) demonstrate stable controllable transport of particles with diameters $d=6.2\!-\!9.5\;$mm (see also Supplemental Movies 1, 22, and 23). 

%In all experiments, the maximum wave amplitude in the three-wave interference region was $|Z|_{\rm max} \simeq 1\!-\!1.5\;$mm.

%FFFFFFFFFFFFFFFFFFFFFFFFFFFFFFFFFFFFFFFFFFFFFFFFFFFFFFFFFFF
\begin{figure}[t]
\centering
\includegraphics[width=\linewidth]{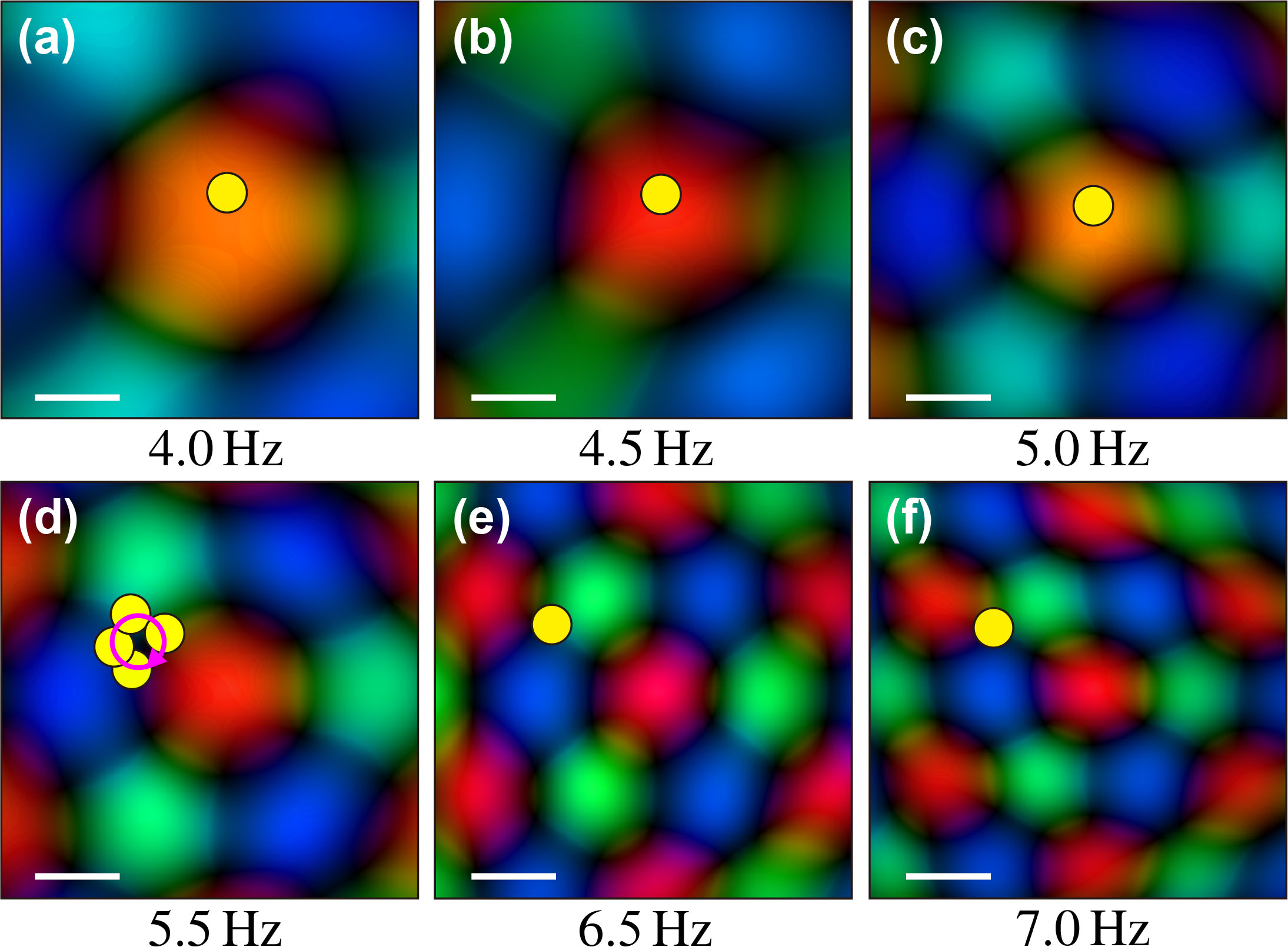}
\caption{Trapping of a particle with diameter $d=9.5\;$mm in monochromatic three-wave interference fields at different frequencies $f=4\!-\!7\;$Hz [the case of $f=6\;$Hz is shown in Fig.~\ref{Fig_exp}(b)]. (a--c) Stable trapping at a wave-intensity maximum. (d) Transition regime with orbital motion around a vortex (field zero). (e,f) Stable trapping at a vortex.}
\label{Fig_trap}
\end{figure}
%FFFFFFFFFFFFFFFFFFFFFFFFFFFFFFFFFFFFFFFFFFFFFFFFFFFFFFFFFFF

%FFFFFFFFFFFFFFFFFFFFFFFFFFFFFFFFFFFFFFFFFFFFFFFFFFFFFFFFFFF
%FFFFFFFFFFFFFFFFFFFFFFFFFFFFFFFFFFFFFFFFFFFFFFFFFFFFFFFFFFF

%%%%%%%%%%%%%%%%%%%%%%%%%%%%%%%%%%%%%%%%%
{\it Concluding remarks.---}
%%%%%%%%%%%%%%%%%%%%%%%%%%%%%%%%%%%%%%%%%
In summary, we have demonstrated stable and controllable 2D transport of floating particles using the interference of water waves with slightly modulated frequencies, which generates a time-varying trapping landscape of structured water waves. Depending on the system parameters, particles can be trapped either at moving wave-intensity maxima or at moving spatiotemporal vortices \cite{Sukhorukov2005, Bliokh2012, Jhajj2016, Chong2020, Zhang2023NC, Che2024PRL, Martin2025NP}. 
%(To the best of our knowledge, this constitutes the first practical application of {spatiotemporal} vortices, which are currently attracting considerable interest.) 
We have explored different trapping regimes and parameter ranges, demonstrating that the transport is robust with respect to variations in particle size and central wave frequency. Importantly, we also implemented the simultaneous transport of two particles trapped at different lattice sites, along either similar or distinct trajectories. {Naturally, the number of independently controllable particle degrees of freedom is limited by the number of independently tunable parameters, which is three in our experiments. Increasing the number of independently controlled wave parameters would provide additional degrees of freedom for manipulation of multiple particles.}

In this work, we employed a triangular lattice formed by the interference of three waves. Alternatively, square lattices generated by the interference of four propagating or two standing waves \cite{Filatov2016_II, Francois2017} can also be used. 

We emphasize that the transport mechanism used here is fundamentally different from time-averaged ponderomotive forces in monochromatic fields \cite{Ashkin_book, Molloy2002CP, Grier2003N, Bustamante2021NR, Gieseler2021AOP, Volpe2023JPP,  Ozcelik2018NM, Meng2019JPD, Baudoin2020AR, Dholakia2020NRP, Toftul2025arxiv} or from analogous Stokes-drift phenomena in fluids \cite{Bremer2018,  Bliokh2022SA}. While such forces are responsible for trapping in our system, the transport itself arises from controlled motion of the trapping landscape \cite{Whitworth1991U, Courtney2012PRSA, Courtney2014APL, Ruffner2012PRL_II, Li2021SA, Morrell2024PRE}. 
%The transport mechanism realized here can be regarded as a 2D water-wave counterpart of ``optical conveyors'' \cite{Ruffner2012PRL_II} or of recently proposed acoustic dynamic trapping landscapes \cite{Morrell2024PRE}. 

Notably, an accurate theoretical description of the water-wave-induced forces governing the nodal and anti-nodal trapping regimes remains a challenging problem {because, unlike in optics and acoustics, the corresponding Mie-like scattering problem for water waves and floating particles has not been elaborated}. Refs.~\cite{Falkovich2005N, Denissenko2006PRL, Lukaschuk2007EPJ} focused on the dependence of the trapping gradient force on the hydrophilic or hydrophobic properties of the particle, whereas here we observed a crucial frequency dependence of the effect, including a transition between the two regimes near the resonant frequency of the particle's oscillations on a still water surface.

Overall, our results advance the water-wave analogs of optical and acoustic structured-wave systems  \cite{Wang2025N, Smirnova2024PRL, Bliokh2022SA, Rozenman2019F, Bush2024APL, Zhu2024NRP, Bliokh2025CP} and establish structured water waves as an efficient reconfigurable platform for wave-based manipulation of particles in the millimeter-to-meter size range, including biological and soft-matter samples. 
Further development of this platform may enable active water-wave-controlled sorting {\cite{Simon2018APL, Kandemir2021JSV}}, assembly, and pulling of floating particles, {as well as the implementation of more complex wavefront-shaping approaches \cite{Orazbayev2024NP}}.

%\vspace{0.5cm}

\begin{acknowledgments}
{\bf Acknowledgments:} This work is supported by
% BW:
the National Natural Science Foundation of China (Grants No.~12522420, 12504440, 12234007, 12321161645, 12221004, T2394480, and T2394481), the National Key R\&D Program of China (Grants No.~2023YFA1406900 and 2022YFA1404800), the China Postdoctoral Science Foundation (Grants No.~2023M741024 and 2024T170218), the Key Research Project of Henan Provincial Higher Education (Grant No.~26A140003),
% KYB:
the European Union (project HORIZON-MSCA-2022-COFUND-01-SmartBRAIN3-101126600),
%the Marie Sk\l{}odowska-Curie COFUND Programme of the European Commission (project HORIZON-MSCA-2022-COFUND-101126600-SmartBRAIN3), 
%ENSEMBLE3 Project (MAB/2020/14) which is carried out within the International Research Agendas Programme (IRAP) of the Foundation for Polish Science co-financed by the European Union under the European Regional Development Fund and Teaming Horizon 2020 programme (GA. No. 857543) of the European Commission, the project of the Minister of Science and Higher Education ``Support for the activities of Centers of Excellence established in Poland under the Horizon 2020 program'' (contract MEiN/2023/DIR/3797), 
% YS:
Singapore Ministry of Education (MOE) AcRF Tier 1 grants (RG157/23 \& RT11/23), Singapore Agency for Science, Technology and Research (A*STAR) MTC Individual Research Grants (M24N7c0080), and Nanyang Assistant Professorship Start Up grant. 

{\bf Data availability:} The data that support the findings of this article are not publicly available upon publication because it is not technically feasible and/or the cost of preparing, depositing, and hosting the data would be prohibitive within the terms of this
research project. The data are available from the authors upon reasonable request.
\end{acknowledgments}

\bibliography{refs}

%\title{End Matter}
%\maketitle

\appendix
%%%%%%%%%%%%%%%%%%%%%%%%%%%%%%%%%%%%%%%%%
\section*{Appendix A}
%%%%%%%%%%%%%%%%%%%%%%%%%%%%%%%%%%%%%%%%%

\setcounter{equation}{0}
\setcounter{figure}{0}
\setcounter{table}{0}
%\setcounter{page}{1}
%\makeatletter
\renewcommand{\theequation}{A\arabic{equation}}
\renewcommand{\thefigure}{A\arabic{figure}}

\begin{figure}[t]
\centering
\includegraphics[width=\linewidth]{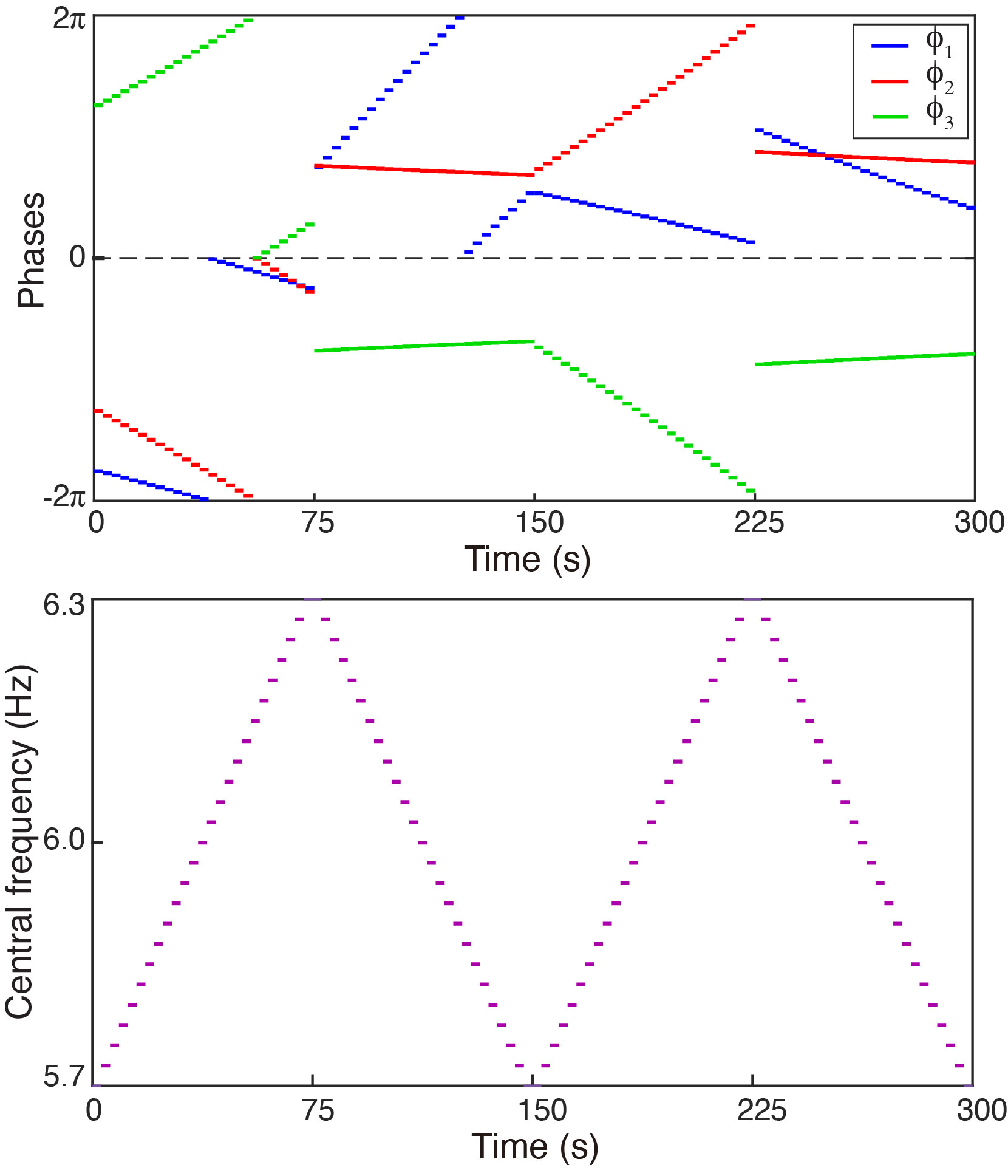}
\caption{Temporal dependencies of the wave phases $\phi_i (t)$ and the central wave frequency $f(t)$ used for the simultaneous manipulation of two particles (Fig.~\ref{Fig_two_particles}).}
\label{Fig_A_A}
\end{figure}

Figure~\ref{Fig_A_A} shows the temporal dependencies of the wave phases, $\phi_i (t)$, and the central wave frequency, $f(t)$, used for the simultaneous manipulation of two particles along square and triangular trajectories, Fig.~\ref{Fig_two_particles}. Linear (on average) variations of the phases $\phi_i (t)$ are equivalent to frequency detunings $\delta\omega_i = d\phi_i/dt$. 
See also Supplemental Movie 13 for the corresponding evolution of the water-wave lattice.

%FFFFFFFFFFFFFFFFFFFFFFFFFFFFFFFFFFFFFFFFFFFFFFFFFFFFFFFFFFF
%FFFFFFFFFFFFFFFFFFFFFFFFFFFFFFFFFFFFFFFFFFFFFFFFFFFFFFFFFFF

\setcounter{equation}{0}
\setcounter{figure}{0}
\setcounter{table}{0}
%\setcounter{page}{1}
%\makeatletter
\renewcommand{\theequation}{B\arabic{equation}}
\renewcommand{\thefigure}{B\arabic{figure}}

%%%%%%%%%%%%%%%%%%%%%%%%%%%%%%%%%%%%%%%%%
\section*{Appendix B}
%%%%%%%%%%%%%%%%%%%%%%%%%%%%%%%%%%%%%%%%%

Here we derive the resonant frequency of the vertical oscillations of a particle floating on a still water surface. Following Ref.~\cite{Lukaschuk2007EPJ}, we consider a spherical particle with density $\rho$, diameter $d$, and mass $m=\pi\rho d^3/6$, submerged in water to a depth $\delta < d$. Neglecting capillary effects, the particle experiences the downward gravitational force $mg$ (where $g$ is the gravitational acceleration) and the upward buoyancy force $-\tilde{m}(\delta)\, g$, where $\tilde{m}(\delta) = \pi \rho_0 (d\delta^2/2 - \delta^3/3)$ is the mass of the displaced water, with $\rho_0$ being the water density. Therefore, the vertical position of the particle obeys the equation of motion
\begin{equation}
\label{eq:EOM}
m\frac{d^2\delta}{dt^2} = \left[ m - \tilde{m}(\delta) \right]g\,.
\end{equation}
The equilibrium position $\delta=\delta_0$ is determined by the condition $\tilde{m}(\delta_0) = m$, while small oscillations about this position, $\delta (t) = \delta_0 + \delta_1 (t)$, are described by the linearized equation 
\begin{equation}
\label{eq:EOM_osc}
m\frac{d^2\delta_1}{dt^2} = - \tilde{m}'(\delta_0) g\, \delta_1\,,
\end{equation}
where $\tilde{m}'(\delta_0) = d\tilde{m}(\delta_0)/d\delta_0 = \pi \rho_0 \delta_0 (d - \delta_0)$.
Hence, $\delta_1$ undergoes harmonic oscillations with frequency
\begin{equation}
\label{eq:frequency}
\omega_0 = \sqrt{\frac{\tilde{m}'(\delta_0) g}{m}} =
\sqrt{\frac{6 g\rho_0 \delta_0 (d - \delta_0)}{\rho d^3}}\,.
\end{equation}
Using the parameters $\rho_0=1\,{\rm g/cm^3}$, $\rho=0.9\,{\rm g/cm^3}$, $g = 980\,{\rm cm/s^2}$, and $d=0.95\,$cm, we calculate the equilibrium submergence depth $\delta_0 \simeq 0.764\,$cm and the corresponding oscillation frequency $f_0 = \omega_0/2\pi \simeq 5.24\,$Hz. This value agrees very well with the transition frequency between the anti-nodal and nodal trapping regimes in Fig.~\ref{Fig_trap}(d).

\end{document}